\documentclass[aps,prb,superscriptaddress,twocolumn]{revtex4-1}
\usepackage{amsmath}
\usepackage{amssymb}
\usepackage{mathtools}
\usepackage{graphicx}
\usepackage[usenames,dvipsnames]{color}
\usepackage{bm}
\usepackage{hyperref}
\usepackage{url}
\usepackage[utf8]{inputenc}
\usepackage{subfigure}
\usepackage{slashed,bbm}
\usepackage{graphics,psfrag,epsfig}
\usepackage{dsfont}
\usepackage{setspace}

\begin{document}

\newcommand{\beq}{\begin{equation}}
\newcommand{\eeq}{\end{equation}}
\newcommand{\beqa}{\begin{eqnarray}}
\newcommand{\eeqa}{\end{eqnarray}}
\newcommand{\note}[1]{{\color{red} [#1]}}
\newcommand{\bra}[1]{\ensuremath{\langle#1|}}
\newcommand{\ket}[1]{\ensuremath{|#1\rangle}}
\newcommand{\bracket}[2]{\ensuremath{\langle#1|#2\rangle}}
\renewcommand{\vec}[1]{\bm{#1}}
\newcommand{\dagga}{{\phantom{\dagger}}}

\newcommand{\txi}{\tilde{\xi}}
\newcommand{\teta}{\tilde{\eta}}

\newcommand{\nn}{\nonumber } 

\newcommand{\todo}[1]{\textcolor{red}{#1}}
\newcommand{\remark}[1]{\textcolor{blue}{#1}}

\title{Bond-ordered states and $f$-wave pairing of spinless fermions on the honeycomb lattice}

\author{S. Hesselmann}
\affiliation{Institut f\"ur Theoretische Festk\"orperphysik, JARA-FIT and JARA-HPC, RWTH Aachen University, 52056 Aachen, Germany}

\author{D. D. Scherer}
\affiliation{Niels Bohr Institute, University of Copenhagen, DK-2100 Copenhagen, Denmark}

\author{M. M. Scherer}
\affiliation{Institute for Theoretical Physics, University of Cologne, 50937 Cologne, Germany}

\author{S. Wessel}
\affiliation{Institut f\"ur Theoretische Festk\"orperphysik, JARA-FIT and JARA-HPC, RWTH Aachen University, 52056 Aachen, Germany}

\date{\today}

\begin{abstract}
Spinless fermions on the honeycomb lattice with repulsive nearest-neighbor interactions are known to harbour a quantum critical point at half-filling, with critical behavior in the Gross-Neveu (chiral Ising) universality class. The
critical interaction strength separates a weak-coupling semimetallic regime from a commensurate charge-density-wave phase. 
The phase diagram of this basic model of correlated fermions on the honeycomb lattice beyond half-filling is, however, less well established. Here, we perform an analysis of its many-body instabilities using the functional renormalization group method with a basic Fermi surface patching scheme, which allows us to treat instabilities in competing channels on equal footing also away from half-filling. 
Between half-filling and  the Van Hove filling, the free Fermi surface is holelike and we again find a charge-density wave instability to be dominant at large interactions. Moreover, its characteristics are those of the half-filled case. Directly at the Van Hove filling, the nesting property of the free Fermi surface stabilizes a dimerized bond-order phase. 
At lower filling, the free Fermi surface becomes electronlike and a superconducting instability with $f$-wave symmetry is found to emerge from the interplay of intra-unit-cell repulsion and collective  fluctuations in the proximity to the charge-density wave instability. We estimate the extent of the various phases and extract the  corresponding order parameters from the effective low-energy Hamiltonians. 
\end{abstract}

\maketitle

\section{Introduction}\label{sec:intro}

Recent experiments on twisted graphene bilayers report evidence for strongly correlated insulating behavior near half-filling with respect to the superlattice unit cell~\cite{Jarillo-Herrero:2018a} and unconventional superconductivity for doping slightly away from it~\cite{Jarillo-Herrero:2018}. While this is undoubtedly a major experimental breakthrough, the physics of correlated electrons in graphene-based systems remains a formidable challenge for theoretical methods, which is often approached in terms of simplified models. A fundamental model to study the behavior of itinerant fermions with interactions in graphenelike systems is the extended Hubbard model for spinless fermions on the honeycomb lattice~\cite{tchougreeff1992charge,PhysRevLett.97.146401,PhysRevB.79.085116}.

In the particle-hole symmetric, half-filled case (corresponding to a vanishing chemical potential $ \mu = 0$), the quantum many-body phase diagram of this model with nearest- and next-to-nearest neighbor interactions has been studied intensely, employing various theoretical approaches ranging from mean-field approximations~\cite{PhysRevLett.100.156401,PhysRevB.81.085105,PhysRevLett.107.106402,PhysRevB.87.085136} over functional renormalization group (fRG) studies~\cite{PhysRevB.92.155137} to various numerical approaches~\cite{PhysRevB.92.085146,PhysRevB.88.245123,PhysRevB.89.035103,PhysRevB.89.165123,PhysRevB.92.085147}. A large variety of competing phases was found, see Ref.~\onlinecite{Capponi:2017} for a review. Most of these theoretical approaches agree on the presence of a stable semimetallic state for small interactions and a competition of different charge-ordered phases and a Kekul\'e phase beyond critical values for the nearest- and next-to-nearest neighbor interaction terms. At the same time, the presence of a topological Chern insulating phase, which was suggested by early mean-field studies~\cite{PhysRevLett.100.156401}, was not generally confirmed. 

Varying only the nearest-neighbor repulsion $V$ at vanishing next-to-nearest neighbor interaction, consensually leads to a continuous quantum phase transition toward a commensurate charge density wave (CDW) state at a critical interaction strength of $V_c\approx 1.36t$, see Refs.~\onlinecite{PhysRevB.93.155117,PhysRevB.93.155157,1367-2630-17-8-085003,1367-2630-16-10-103008}. Here, $t$ denotes the nearest-neighbor tight-binding hopping amplitude.  The quantum critical behavior of the semimetal-to-CDW transition of the spinless fermions is suggested to belong to  the three-dimensional Gross-Neveu universality class with an Ising order parameter~\cite{PhysRevLett.97.146401,PhysRevB.80.075432}, see Refs.~\onlinecite{PhysRevD.96.114502,Iliesiu:2017nrv,PhysRevB.94.245102,Gracey:2016mio,PhysRevD.96.096010} for recent estimates from different theoretical methods.
Beyond half-filling (i.e., at finite chemical potential $\mu \neq 0$) the physics of spinless fermions on the honeycomb lattice is far less explored; see, for example, the mean-field studies in Refs.~\onlinecite{PhysRevLett.107.106402,PhysRevB.87.085136} and the numerical approaches of Refs.~\onlinecite{PhysRevB.94.075144, Broecker2017}. In fact, a thorough study of the system away from half-filling is hampered by the presence of a sign-problem for lattice quantum Monte Carlo simulations at finite chemical potential and the high numerical costs related with some otherwise promising methods such as, e.g., exact diagonalization.

Another theoretical approach, which has proven itself to be  promising for an unbiased identification of the leading quantum many-body instabilities of correlated lattice fermion systems is the fermionic fRG~\cite{RevModPhys.84.299,doi:10.1080/00018732.2013.862020}. In the context of spinless fermions on the honeycomb lattice with nearest- and next-to-nearest neighbor interactions it has provided support for the suppression of a topological Chern insulating phase and a competition among various charge-modulated phases~\cite{PhysRevB.92.155137}. Therefore, and despite differences at larger couplings, the fRG is qualitatively in line with  numerically exact approaches and allows for the inclusion of correlations beyond mean-field theory.
Moreover, the fRG can straightforwardly be extended to finite chemical potentials, which has already been explored in the case of spin-1/2 fermions on the honeycomb lattice, both at small doping~\cite{PhysRevLett.100.146404,PhysRevB.87.094521} as well as near the Van Hove filling, to reveal the possibility of unconventional superconductivity~\cite{PhysRevB.86.020507,PhysRevB.90.045135,PhysRevB.85.035414}. A corresponding study for the simpler case of the paradigmatic spinless fermion model, however, is still lacking. Here, we take the recent experiments~\cite{Jarillo-Herrero:2018a,Jarillo-Herrero:2018} on graphene superlattices as a further motivation to fill this gap. We note that in combination with complementary theoretical approaches, the study of this minimal model at finite doping can be expected to constitute one of the most fundamental building blocks, when it comes to the understanding of correlated phases and unconventional superconductivity in graphene.

\begin{figure}[ht]
\includegraphics[width=0.9\columnwidth]{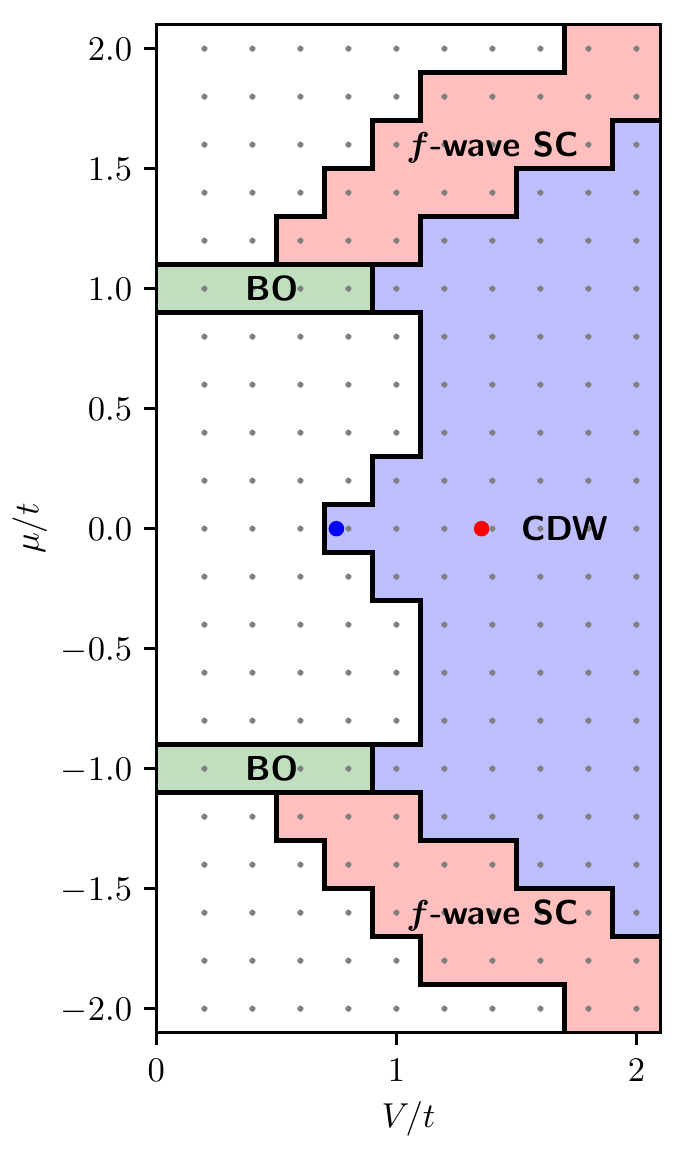}
\caption{(Color online) Tentative phase diagram of spinless fermions on the honeycomb lattice with nearest-neighbor repulsion $V$ and chemical potential $\mu$, given in units of the nearest-neighbor hopping amplitude $t$. At $\mu=0$, the transition from  the semi-metallic to the commensurate CDW phase in mean-field theory is given by $V_c^{\text{MF}} \approx 0.75t$ (blue dot) and from quantum Monte Carlo simulations by $V_c^{\text{QMC}} \approx 1.36t$ (red dot).
For this phase diagram, we have evaluated the flow equations with the patching schemes from Fig.~\ref{fig:patch}, resolving instabilities down to scale of  
$\Lambda/t = 10^{-7}$, and we have chosen a wave vector resolution of $N=120$ patches in the Brillouin zone. The initial parameters of the flow are indicated by the grey dots.}
\label{fig:phase_diagram}
\end{figure}

In the following, we will present results from a thorough fRG study for the quantum many-body instabilities of spinless fermions on the honeycomb lattice with nearest-neighbor repulsion at finite chemical potential. As our main result, we present the tentative phase diagram as a function of interaction strength and chemical potential in Fig.~\ref{fig:phase_diagram}. In addition to the instability toward the commensurate CDW state (which is already well established at $\mu=0$), we identify an instability toward a bond-ordered (BO) regime corresponding to  the Van Hove singularity point. In the low-density regime, we furthermore find a charge-fluctuation driven instability in the Cooper channel, leading to an $f$-wave superconducting (SC)  state. The details of this phase diagram will be discussed in the following sections. 

%
%

The remainder of this paper is organized as follows. In Sec.~\ref{sec:model}, we introduce the model Hamiltonian for spinless fermions on the honeycomb lattice with repulsive interaction at finite chemical potential. In Sec.~\ref{sec:method}, we review the fRG method utilized to carry out the instability analysis, and discuss the truncation scheme and approximations employed. In Sec.~\ref{sec:instability}, we then present an instability analysis within the various regimes in the parameter space of the model Hamiltonian. The results of this analysis are then combined in the tentative phase diagram, cf. Fig.~\ref{fig:phase_diagram}, as a function of the nearest-neighbor repulsion strength and the chemical potential. 
We conclude in Sec.~\ref{sec:conclusions}, along with a further discussion of our findings. 

\section{Model}~\label{sec:model}

\begin{figure}[ht]
\includegraphics[width=0.9\columnwidth]{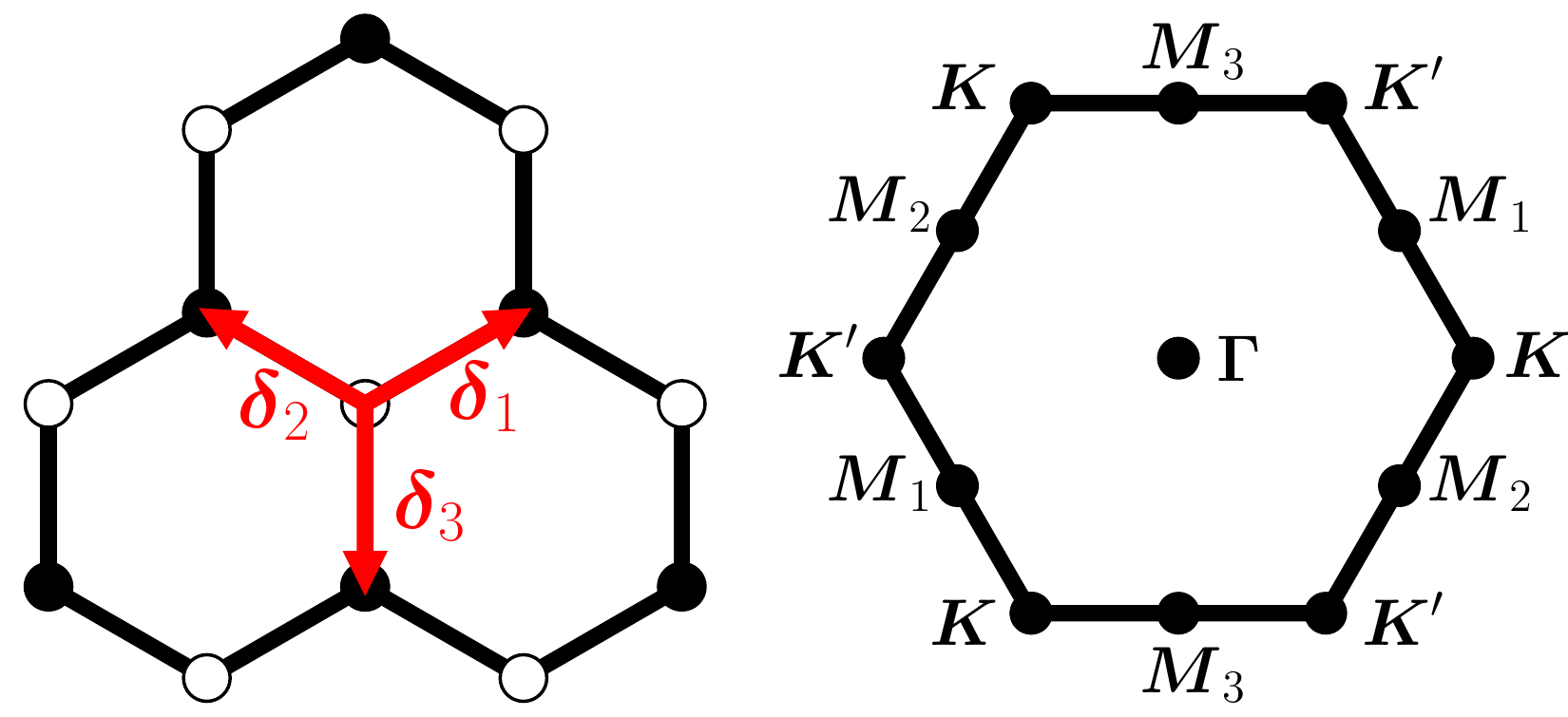}
\caption{(Color online) Left panel:  honeycomb lattice with the two sublattices $A$ and $B$ indicated  by  white and black circles, respectively. Next-nearest neighbors are connected by the three vectors $\bm{\delta}_i$, $i=1,2,3$. Right panel: Brillouin zone, with the nonequivalent high-symmetry points  $\bm{\Gamma}$, $\bm{K}$, and $\bm{K}^\prime$, as well as $\bm{M}_1$, $\bm{M}_2$, and $\bm{M}_3$ indicated as labeled.}
\label{fig:lattice}
\end{figure}

We consider spinless fermions on a honeycomb lattice with nearest-neighbor interactions, also known as the $t-V$ model, and include  a chemical potential term to control the filling. The full Hamiltonian thus  consists of two parts $H = H_0 + H_{\text{int}}$. 
The tight-binding part of the Hamiltonian, $H_0$, can be further decomposed into nearest-neighbor hopping terms and the on-site chemical potential term,
\begin{align}\label{eq:tb}
H_0 &= -t \sum_{\langle \bm{i}, \bm{j} \rangle} \left( c_{\bm{i}}^\dag c_{\bm{j}} + c_{\bm{j}}^\dag c_{\bm{i}} \right) - \mu \sum_{\bm{i}} c_{\bm{i}}^{\dag} c_{\bm{i}},
\end{align}
where the operators $c_{\bm{i}}^\dag$ and $c_{\bm{i}}$ create and destroy, respectively, a spinless fermion at the lattice site ${\bm i}$, $t$ is the nearest-neighbor hopping amplitude and $\mu$ the chemical potential. The sum $ \langle \bm{i}, \bm{j} \rangle $ is restricted to nearest neighbors on the honeycomb lattice. We can express $H_0$ in the orbital basis by performing a Fourier transform, defined by
\begin{align}
c_{\bm{k},A} = \frac{1}{\sqrt{\mathcal{N}}}\sum_{\bm i \in A} \mathrm{e}^{\mathrm{i} {\bm k}\cdot{\bm i} } c_{\bm{i}}, \quad
c_{\bm{k},B} = \frac{1}{\sqrt{\mathcal{N}}}\sum_{\bm i \in B} \mathrm{e}^{\mathrm{i} {\bm k}\cdot{\bm i} } c_{\bm{i}},
\end{align}
with $\mathcal{N}$ the number of two-atom unit cells and the corresponding transformations for $ c_{\bm{k},A/B}^{\dagger} $ such that
\begin{align}
H_0 = \sum_{\bm{k}} \begin{pmatrix} c_{\bm{k}, A}^\dag & c_{\bm{k}, B}^\dag \end{pmatrix} \hat{h} (\bm{k}) \begin{pmatrix} c_{\bm{k}, A} \\ c_{\bm{k}, B} \end{pmatrix},
\end{align}
with the Bloch Hamiltonian 
\begin{align} 
\hat{h} (\bm{k}) = 
- \begin{pmatrix} 
\mu & t d_{\bm k}^{} \\
t d_{\bm k}^{\ast} & \mu 
 \end{pmatrix},
\end{align}
where $ d_{\bm k} = \sum_{i} \mathrm{e}^{\mathrm{i} {\bm k} \cdot {\bm \delta}_{i} } $ and $ {\bm \delta}_{i} $, $ i = 1,2,3 $ are the three nearest-neighbor vectors pointing from the $A$ to the $B$ sublattice as shown in Fig.~\ref{fig:lattice}. The operators $c_{\bm{k}, o}, c_{\bm{k}, o}^\dag$ correspond to single-particle basis-states with Bloch momentum $\bm{k}$ and orbital (i.e., sublattice) index $o \in \{A, B\}$. 
In the following, we also employ the notation $\bar{o}=B,A$ to denote the other sublattice for   $o=A,B$, respectively. Diagonalization of the free Hamiltonian by a unitary transformation $ c_{{\bm k},o} = \sum_{b} u_{o,b}({\bm k}) c_{{\bm k},b}$ leads to 
\begin{align}
H_0 = \sum_{\bm{k}} \sum_{b} \left( \epsilon_b(\bm{k}) - \mu \right) c_{\bm{k}, b}^{\dag} c_{\bm{k}, b}\,,
\end{align}
where $\epsilon_b(\bm{k})$ is the tight-binding energy dispersion of free fermions on the honeycomb lattice, which features two energy bands $b \in \{+,-\}$ with the two inequivalent characteristic Dirac points at the $K$ and $K^\prime$ points in the first Brillouin zone (BZ), respectively. In the following, we will use this band basis to set up the fRG.

%
%

The interacting part of the Hamiltonian, $H_{\text{int}}$, contains density-density interactions between all sites along the  nearest-neighbor bonds,
\begin{align}\label{eq:ia}
H_{\text{int}} &= V \sum_{\langle \bm{i}, \bm{j} \rangle} \left( c_{\bm{i}}^\dag c_{\bm{i}} - \frac{1}{2} \right) \left( c_{\bm{j}}^\dag c_{\bm{j}} - \frac{1}{2} \right)\,.
\end{align}
At half-filling, it is well-established that $H_{\text{int}}$ drives 
a continuous quantum phase transition toward a commensurate CDW phase when the interaction strength $V$ exceeds the critical value~\cite{PhysRevB.93.155117,PhysRevB.93.155157,1367-2630-17-8-085003,1367-2630-16-10-103008} $V_c/t\approx 1.36(1)$.
In the following, we are interested in exploring the phase diagram beyond the half-filled case. 
Due to the particle-hole symmetry of the Hamiltonian $H$ about half-filling, we can constrain the fRG analysis to the regime with a negative chemical potential, $\mu<0$, which corresponds to fermion densities below half-filling. The case of larger fillings for positive $\mu>0$ can  be obtained by a particle-hole transformation. It also follows directly from the analysis of the single particle (hole) problem, that the lattice is empty (full) for $\mu<-3/2V-3t$ ($\mu>3/2V+3t$).  

For the following fRG analysis, Fourier transformation of creation and annihilation operators and subsequently applying the unitary transformation $u_{o,b}({\bm k})$ allows to represent the interacting part $H_{\text{int}}$ in terms of the basis of single-particle eigenstates of $H_{0}$. While the Fourier-transform yields a momentum dependent interaction due to the nearest-neighbor repulsion, the matrix elements $u_{o,b}({\bm k})$ imprint an additional wave-vector dependence (often called "orbital make-up") on the interaction, as the band-basis states are obtained by a ${\bm k}$-dependent hybridization from the $A$ and $B$ sublattice Bloch states. In fact, while $  | u_{A,\pm}(\vec{k})  | =  | u_{B,\pm} (\vec{k}) | = 1/\sqrt{2}$, the $A$ and $B$ components for a given band differ by a $ \vec{k}$-dependent phase. We can then compactly express the interaction Hamiltonian in the band basis as
\begin{align}\label{eq:ib}
H_{\text{int}} = \frac{1}{4} \sum_{\{b_i\}} \sum_{\{\bm{k}_i\}} & V_{b_1, b_2, b_3, b_4} (\bm{k}_1, \bm{k}_2, \bm{k}_3, \bm{k}_4) \times 
\\ 
& c_{\bm{k}_1, b_1}^{\dag} c_{\bm{k}_2, b_2}^{\dag} c_{\bm{k}_3, b_3} c_{\bm{k}_4, b_4}\,, \nonumber
\end{align}
where $V_{b_1, b_2, b_3, b_4}$ includes antisymmetric combinations of the $\bm{k}$-dependent interaction and we have absorbed single-particle terms in Eq.~\eqref{eq:ia} into a chemical potential term. We note that a momentum-conserving $\delta$-function is implicit in Eq.~\eqref{eq:ib}.

\section{Method}~\label{sec:method}

In this paper, we investigate the instabilities of the model defined by the Hamiltonian $H = H_{0} + H_{\mathrm{int}}$. Such instabilities indicate ordering tendencies of the quantum many-body ground state. To that end, we employ the fRG approach~\cite{Wetterich:1992yh,RevModPhys.84.299,doi:10.1080/00018732.2013.862020} for the one-particle irreducible vertex function with an energy cutoff. In this scheme, the bare propagator is modified by an infrared regulator at an energy scale $\Lambda$. The renormalization-group flow is generated by successively integrating out fermionic degrees of freedom from an initial scale $\Lambda_0 \sim 3 t$ down toward $\Lambda \rightarrow 0$. The fRG approach allows for an unbiased identification of the leading instability, as generated during the flow in the presence of competing fluctuations in other channels. In the following, we briefly review the basic fRG formulation and discuss the usual approximations used in practical calculations. 

The fRG flow equations are most conveniently derived by switching from the Hamiltonian formulation to an action-based formulation  of the quantum many-body system. The starting point is therefore the fermionic imaginary-time action,
\begin{align}
\mathcal{S} [ \Psi, \bar{\Psi} ] = - \left(\bar{\Psi}, \mathcal{G}_0^{-1} \Psi \right) + \mathcal{V}[ \Psi, \bar{\Psi} ],
\end{align}
where $\Psi(\xi), \bar{\Psi}(\xi)$ are Grassmann fields, $\mathcal{G}_0(\xi,\xi^{\prime}) $ is the propagator of the non-interacting system, and we defined the multi-index $ \xi = (\omega_{n},{\bm k},b) $ for compact notation. 
The quadratic part reads
\begin{align}
\left(\bar{\Psi}, \mathcal{G}_0^{-1} \Psi \right) = \sum_{\xi,\xi^{\prime}}  
\bar{\Psi}(\xi) [\mathcal{G}_0^{-1}](\xi,\xi^{\prime}) \Psi(\xi).
\end{align}
In the band basis, the propagator is diagonal with respect to $ \xi $ and can be expressed as
\begin{align}
\mathcal{G}_0(\xi,\xi^{\prime}) = \mathcal{G}_0(\omega_n, \bm{k}, b) \delta_{\xi,\xi^{\prime}},
\end{align}
with 
\begin{align} 
\mathcal{G}_0(\omega_n, \bm{k}, b) = \frac{1}{i \omega_n - \epsilon_b(\bm{k}) + \mu}.
\end{align}
Further, $\mathcal{V}$ is a quartic interaction functional,
\begin{align}
\mathcal{V}[ \Psi, \bar{\Psi} ] = \frac{1}{4} \sum_{\{\xi_i \}} V(\xi_1, \xi_2, \xi_3, \xi_4)
\bar{\Psi}(\xi_{1}) \bar{\Psi}(\xi_{2}) \Psi(\xi_3) \Psi(\xi_4), \nonumber
\end{align}
and the bare interaction vertex in the band picture $ V(\xi_1, \xi_2, \xi_3, \xi_4) $ is obtained from Eq.~\eqref{eq:ib} by multiplying
by a $\delta$-function for Matsubara frequencies due to the static nature of the microscopic interaction
Hamiltonian. The generating functional for connected correlation functions is then obtained by performing 
the functional integral over Grassmann fields in the presence of Grassmann-valued source fields $\bar{\eta}$, $\eta$,
\begin{align}
G[\eta,\bar{\eta}] = -\ln\int\!\mathcal{D}[\bar{\Psi},\Psi] \mathrm{e}^{-\mathcal{S}[\Psi,\bar{\Psi}] + (\bar{\eta},\Psi) + (\bar{\Psi},\eta)}.
\end{align}
Taking the Legendre transform with respect to the source fields, we obtain the generating functional $\Gamma$ for one-particle irreducible (1-PI) vertices, 
\begin{align}
\Gamma[\phi,\bar{\phi}] = (\bar{\eta},\phi) + (\bar{\phi},\eta) + G[\eta,\bar{\eta}],
\end{align}
where $ \phi = -\frac{\delta}{\delta \bar{\eta}} G[\eta,\bar{\eta}] $, $ \bar{\phi} = \frac{\delta}{\delta \eta} G[\eta,\bar{\eta}] $ and we have for simplicity suppressed indices in the notation of the field variables. The 1-PI vertices $\Gamma^{(2n)}$ appear as the coefficients of an expansion of $ \Gamma[\phi,\bar{\phi}] $ in the fields $ \phi $, $ \bar{\phi}$. This is the typical point of departure to derive a set of equations of motion governing the behavior of the vertex functions, which then have to be solved in an appropriate approximation. Within the fRG, however, the governing equations are obtained by recasting the fermionic functional integral in a scale-dependent way.

An energy cutoff in the propagator is introduced such that the bare propagator becomes
\begin{align}
\mathcal{G}_0(\omega_n, \bm{k}, b) \rightarrow \mathcal{G}_0^{\Lambda}(\omega_n, \bm{k}, b) = \frac{\Theta_{\epsilon}(\epsilon_b(\bm{k}) - \mu)}{i \omega_n - \epsilon_b(\bm{k}) + \mu},
\end{align}
where $\Theta_{\epsilon}$ is taken as a smoothed Fermi function that cuts off modes with energies $| \epsilon_b(\bm{k}) - \mu | \lesssim \Lambda$. The regularized propagator can be used to obtain the $\Lambda$-dependent functional integral for the effective action $\Gamma^{\Lambda}[\phi,\bar{\phi}]$, which now generates the scale-dependent 1-PI vertex functions $\Gamma^{(2n),\Lambda}$. The vertex functions $\Gamma^{(2n),\Lambda}$ are the central objects that are monitored in the course of the renormalization-group flow upon the successive integration of fermionic modes. Taking the variation with respect to $\Lambda$ generates a hierarchy of coupled flow equations for $\Gamma^{(2n),\Lambda}$, which continuously connect the bare vertices $\Gamma^{(2n),\Lambda_0}$ to the full effective effective vertices that emerge as $\Lambda \rightarrow 0$.
%
\begin{figure}[t!]
\centering
\includegraphics[width=\columnwidth]{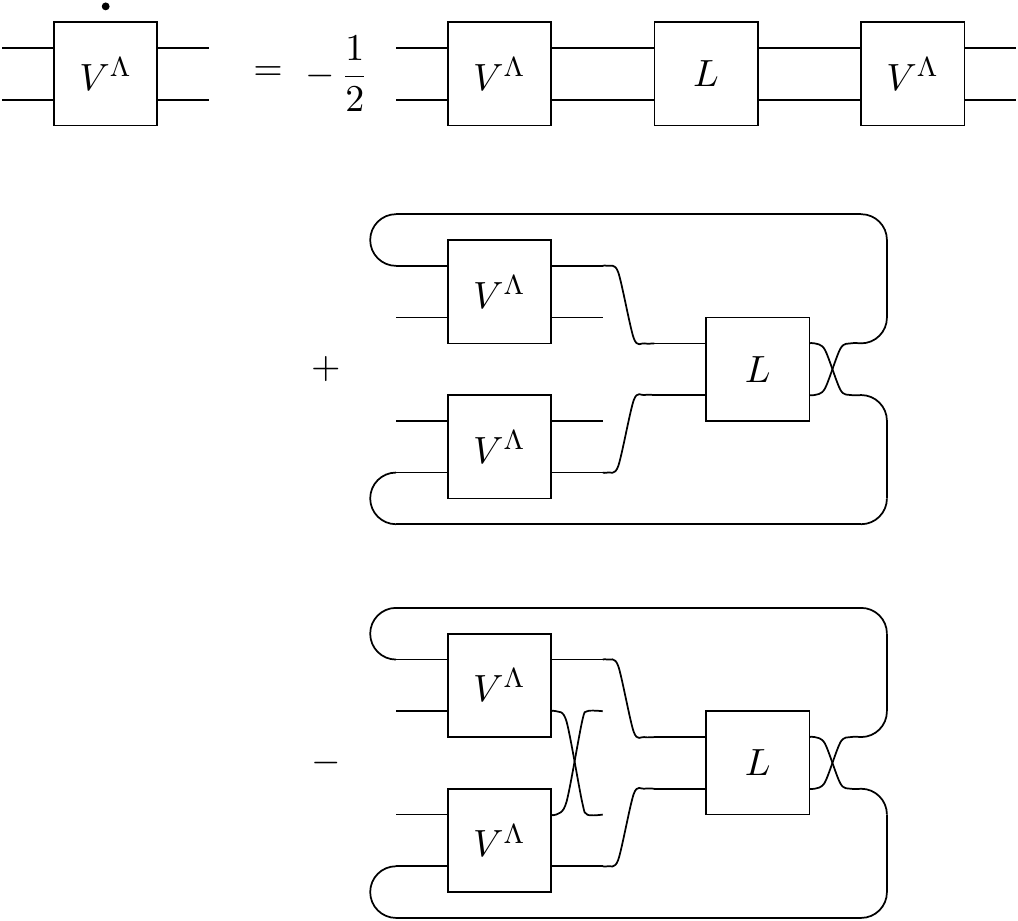}
\caption{Pictorial representation of the fRG flow equation
for the four-point vertex function $V^{\Lambda}$ using Feynman diagrams. The boxes represent $V^{\Lambda}$ and the loop kernel $L$, while the
black dot on the left hand side indicates the scale-derivative $d/d\Lambda V^{\Lambda}$. Internal variables are contracted according to the connected lines, see Eq. \eqref{eq:flow_pp} for the algebraic expression. The first diagram on the right hand side represents the particle-particle channel, the other two show the direct and the crossed particle-hole channel, respectively. A detailed derivation of the diagrammatic rules can be found in Ref.~\onlinecite{PhysRevB.93.115111}.}
\label{fig:diagrams}
\end{figure}
%
In practice, one has to truncate the full hierarchy of flow equations to make a numerical integration feasible. Here, we follow the approach from Ref.~\onlinecite{Salmhofer01012001}, focusing on the flow of the $4$-point vertex function $\Gamma^{(4),\Lambda}$, neglecting the feedback of the flowing self-energy $\Sigma^{\Lambda}=\Gamma^{(2),\Lambda}$ as well as higher order vertex functions $\Gamma^{(2n),\Lambda}$ with $n \geq 3$. During the flow, we monitor for divergences of the vertex function, which  indicate divergent physical susceptibilities and therefore signal phase transitions through a flow to strong coupling. The flow is terminated near the critical energy scale $\Lambda_c>0$, and one then examines the divergent vertex structure, which encodes information about the emerging symmetry-broken state of the many-body system. This truncation scheme has been shown to allow for the competition of different fluctuation channels that drive phase transitions, and has been used successfully to study instabilities in various two-dimensional fermion systems~\cite{RevModPhys.84.299,doi:10.1080/00018732.2013.862020}.

Within this truncation scheme, the resulting flow equation for the four-point vertex function $V^{\Lambda} \equiv \Gamma^{(4), \Lambda}$ is given by
\begin{align}
\frac{d}{d \Lambda} V^{\Lambda} = \Phi_{\text{pp}}^{\Lambda} + \Phi_{\text{ph,d}}^{\Lambda} + \Phi_{\text{ph,cr}}^{\Lambda}, \label{eq:flow_eq}
\end{align}
where the contributions to the right hand side are given by the particle-particle bubble $\Phi_{\text{pp}}$ as well as the direct and crossed particle-hole bubbles $\Phi_{\text{ph, d}}$ and $\Phi_{\text{ph, cr}}$\textemdash see Fig.~\ref{fig:diagrams} for a diagrammatic representation of the flow equation in Eq.~\eqref{eq:flow_eq}. More explicitly, the loop contributions for the particle-particle and the direct particle-hole channel are given by
\begin{align}
\Phi_{\mathrm{pp}}(\xi_1,\xi_2,\xi_3,&\xi_4) = -\frac{1}{2}\prod_{\nu=1}^{4}\!\int\!\!d \eta_{\nu}\, L(\eta_2,\eta_1,\eta_3,\eta_4) \nn  \\
&\times V^{\Lambda}(\xi_2,\xi_1,\eta_2,\eta_3)V^{\Lambda}(\eta_4,\eta_1,\xi_3,\xi_4), \\ \label{eq:flow_pp}
\Phi_{\mathrm{ph,d}}(\xi_1,\xi_2,\xi_3,&\xi_4) = \prod_{\nu=1}^{4}\!\int\!\!d \eta_{\nu}\, L(\eta_1,\eta_2,\eta_3,\eta_4) \nn \\
&\times V^{\Lambda}(\eta_4,\xi_2,\xi_3,\eta_1)V^{\Lambda}(\xi_1,\eta_2,\eta_3,\xi_4),
\end{align}
and the crossed particle-hole contribution is given through
\begin{align}
\Phi_{\mathrm{ph,cr}}(\xi_1,\xi_2,\xi_3,\xi_4)  =  -\Phi_{\mathrm{ph,d}}(\xi_1,\xi_2,\xi_4,\xi_3)\,,
\end{align}
where the loop kernel $L = S^{\Lambda} \mathcal{G}_{0}^{\Lambda} + \mathcal{G}_{0}^{\Lambda} S^{\Lambda}$ is composed out of the bare scale-dependent propagator $\mathcal{G}_{0}^{\Lambda}$ and the single-scale propagator $S^{\Lambda} = - d/d\Lambda\, \mathcal{G}_{0}^{\Lambda}$, cf. Ref.~\onlinecite{Salmhofer01012001}. Furthermore, we introduced the shorthand notation $\int\!\!d \eta$ for the integration/summation over the various loop variables. For simplicity, we focus on the static part of the vertex function $V^{\Lambda}(\xi_{1},\xi_{2},\xi_{3},\xi_{4})|_{\{\omega_i=0\}}$, which is expected to provide the most singular contribution at the critical scale~\cite{Salmhofer01012001}.

\begin{figure}[t]
\includegraphics[width=1.0\columnwidth]{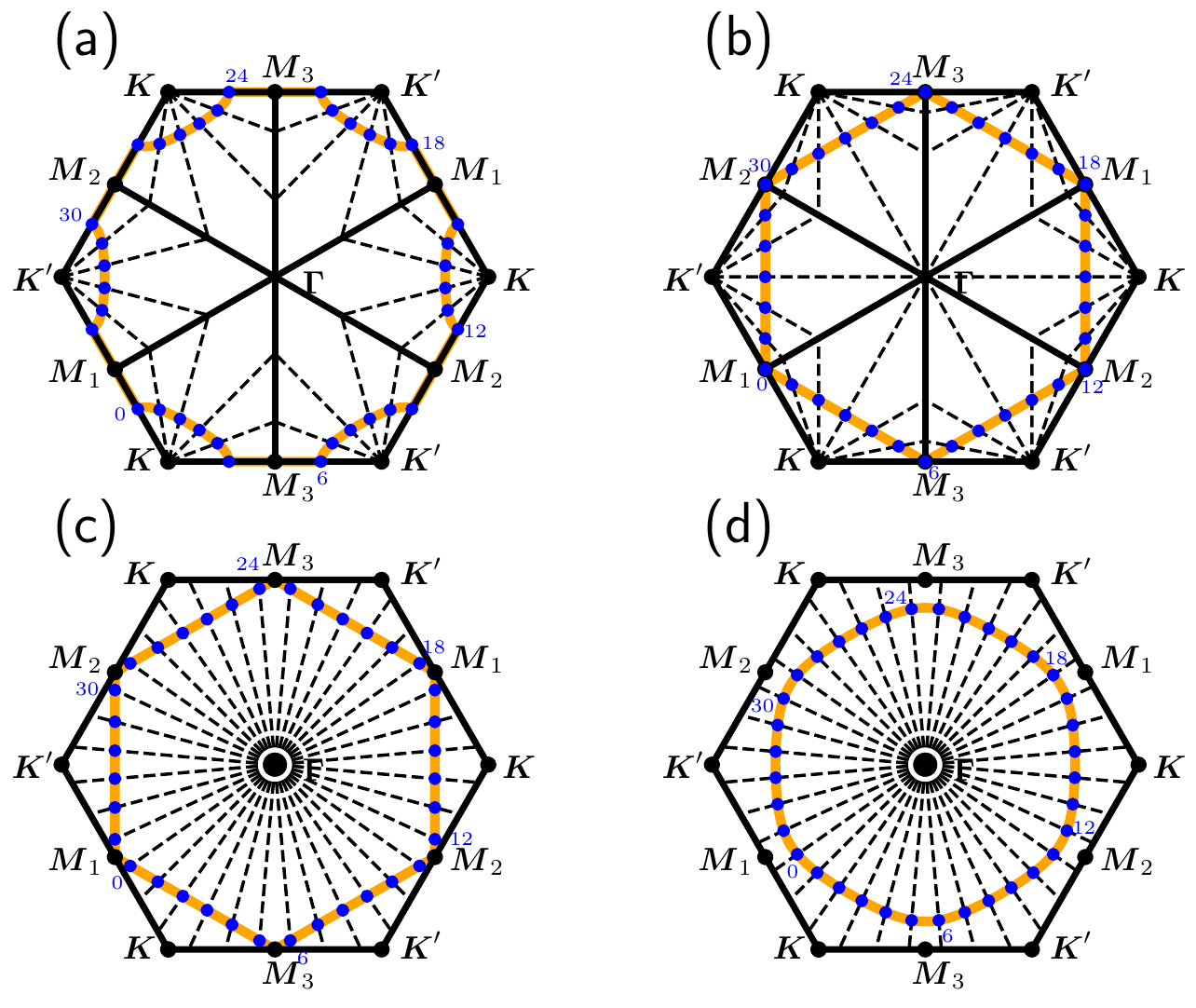}
\caption{(Color online) Momentum discretization schemes of the BZ for different values of the chemical potential, $\mu/t=-0.8$ (a),  $\mu/t=-1$ (b,c), $\mu/t=-1.2$ (d). For better visualization, the discretization scheme is plotted here only for $N=36$ patch points, while the actual calculations were performed with up to $N=120$ patch points. The free Fermi surface is indicated by the orange line, and the blue points represent the projected momentum patch points. They are enumerated by the index function $\pi(\bm{k})$ in the order denoted by the blue numbers. The dashed line indicates the directions along which the numerical integration of the loop kernels are performed. In panels (a) and (b), these originate from the $K$ and $K^\prime$ points and in panels~(c) and~(d) from the $\Gamma$ point.}
\label{fig:patch}
\end{figure}

We solve Eq.~\eqref{eq:flow_eq} by numerical integration, for which we discretize the wave-vector dependence of the vertex function $V^{\Lambda}(\xi_1, \xi_2, \xi_3, \xi_4)$.  The discretization of the momentum dependence is given in terms of a complete patching scheme of the BZ, which projects the momenta $\bm{k}_1, \bm{k}_2$, and $\bm{k}_3$ onto points along the Fermi surface. In this way, we can associate to  each momentum $\bm{k}$ from the BZ an integer index $\pi(\bm{k})$ that extends from $0$ to  $N-1$, where $N$ denotes the number of patches,  cf.~Fig.~\ref{fig:patch}. The fourth momentum $\bm{k}_4$ is given by momentum conservation and is projected onto the closest momentum patch  $\pi(\bm{k}_4)$. The topology of the Fermi surface depends on the chemical potential $\mu$ and changes at the Van Hove singularity point ($|\mu| = t$) from being $K$-point centered to $\Gamma$-point centered. We therefore also change the evaluation of the loop kernels from being $K$-point centered for $|\mu| < t$ to $\Gamma$-point centered for $|\mu| > t$. Right at the Van Hove singularity point,  both discretization schemes can be used and provide consistent results. A pictorial representation of the patching scheme for different $\mu$ is provided in Fig. \ref{fig:patch}. For the actual calculations, we used up to $N=120$ patches.

%
%

The integration of the flow equation is most conveniently performed in the band basis. However, for the physical interpretation of the final vertex structure it can be beneficial to revert to the original orbital basis denoted by the sublattice indices $o=A,B$ by applying the inverse of the unitary transformation $ u_{o,b}({\bm k}) $. Suppressing $\delta$-functions in the notation, we thereby obtain 
\begin{align}
V_{o_{1}o_{2}o_{3}o_{4}}^{\Lambda}({\bm k_{1}},{\bm k_{2}},{\bm k_{3}},{\bm k_{4}})  =\hspace{-0.3cm} \sum_{\{b_{i}\}, \{ {\bm k}_{i} \}} \hspace{-0.3cm} V^{\Lambda}(\xi_{1},\xi_{2},\xi_{3},\xi_{4})|_{\{\omega_i=0\}} \nonumber \\
\times u_{o_{1},b_{1}}({\bm k}_{1}) u_{o_{2},b_{2}}({\bm k}_{2})
u_{o_{3},b_{3}}^{\ast}({\bm k}_{3}) u_{o_{4},b_{4}}^{\ast}({\bm k}_{4})
& &\nonumber
\end{align}
as the central element of our analysis. We therefore perform a transformation back to the orbital basis after the termination of the flow.

Finally, it should be noted that our approach corresponds to a treatment in the grand canonical ensemble, which is appropriate in systems for which phase separation may emerge at strong interactions. As we will discuss below, this is likely the case here. Furthermore, since in the truncated flow equations self-energy feedback is absent, the chemical potential $\mu$ does not get adjusted to a constant filling, and instead it  defines the filling in terms of the initial bare system.

\section{Instability Analysis}~\label{sec:instability}

To evaluate the flow according to Eq.~\eqref{eq:flow_eq}, we use an initial condition $V^{\Lambda_0}$ at the ultraviolet scale $\Lambda_0$, which is determined from the bare interaction term, Eq.~\eqref{eq:ib}. The flow equations are then integrated numerically by successively lowering the energy cutoff scale $\Lambda$. When the initial interaction $V^{\Lambda_0}$ is sufficiently large, some components of the interaction vertex $V^{\Lambda}$ may  increase strongly during the course of  the flow and develop a singularity at a critical scale $\Lambda_c>0$. This is indicative of a quantum many-body instability and suggests a transition toward a symmetry-broken ground state. In practice, the flow thus has to be stopped before it reaches $\Lambda_c>0$, i.e., at a scale $\Lambda^\ast > \Lambda_c$. Close to this critical scale, the effective interaction vertex develops a pronounced momentum structure, which can be used to extract an effective low-energy Hamiltonian and  identify the leading order parameter. In the following, we present the results of the instability analysis separately for three different regimes, according to the value of the chemical potential.


\subsection{Charge-density-wave instability}

%
\begin{figure}[t]
\includegraphics[width=\columnwidth]{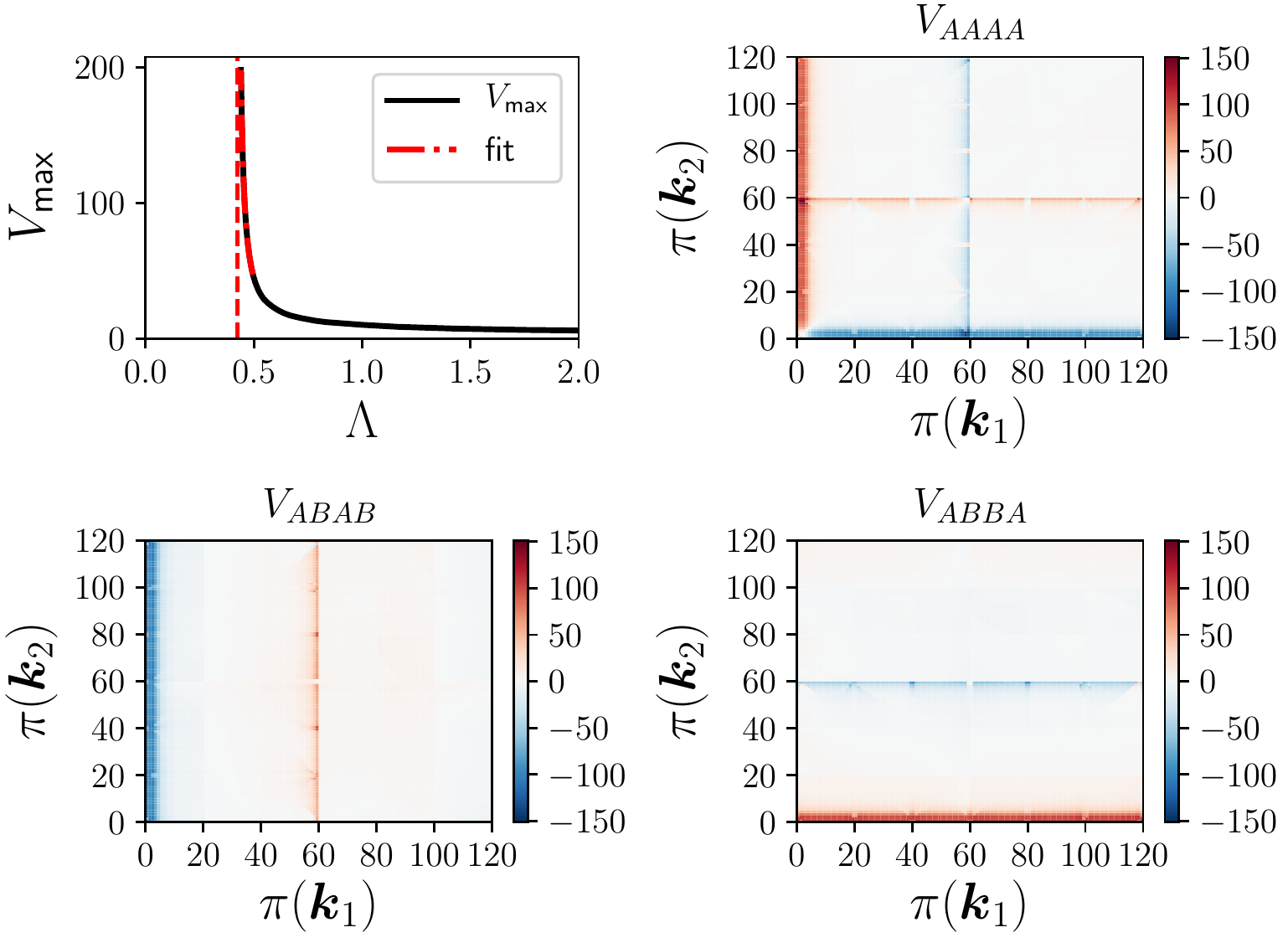}
\caption{(Color online) Upper left panel: Largest vertex component $V_{\mathrm{max}}$ as a function of the RG scale $\Lambda$ for $\mu/t=-0.6$ and $V/t=1.8$. We identify the critical scale at $\Lambda_c/t \approx 0.424$. Upper right panel and bottom panels: Vertex structure for the divergent CDW correlations for the sublattice combinations $V_{AAAA}^{\Lambda_{\mathrm{c}}}$,  $V_{ABAB}^{\Lambda_{c}}$, and $V_{ABBA}^{\Lambda_{c}}$, respectively. Here, we have chosen $N=120$ patch points following the scheme indicated in Fig.~\ref{fig:patch}~(a) and $\pi(\bm{k}_{3})$ is fixed to the first patch point.  The divergent wave-vector structure can be translated to the effective Hamiltonian in Eq.~\eqref{eq:HCDW}.}
\label{fig:vertex_cdw}
\end{figure}
%

We start the instability analysis for the case in which the chemical potential lies within the interval $-1 <\mu/t < 0$.
For the non-interacting system, this corresponds to the density regime between half-filling ($\mu=0$) and the Van Hove singularity point ($\mu/t=-1$), cf. Figs.~\ref{fig:patch}(a)--\ref{fig:patch}(c). Here, we identify an instability of the metallic phase beyond a critical value for the nearest-neighbor repulsion $V_c(\mu)$, which depends on the value of $\mu$. In the following, we concentrate on the real part of the vertex function, and note that the imaginary part develops either subleading instabilities or vanishes completely. The diverging wave-vector structure on the patch points is shown in Fig.~\ref{fig:vertex_cdw} for the case of $\mu/t=-0.6$. The relevant features extracted from the vertex structure are (i)  a vanishing momentum transfer and (ii) a momentum-independent structure factor, which corresponds to a CDW instability, cf. also Ref.~\onlinecite{PhysRevB.92.155137}. 
Note that the feature for $\pi(\bm{k}_1) \approx 60$ and $\pi(\bm{k}_2) \approx 60$ does not correspond to any further finite momentum transfer. 
From this analysis, we thus extract the effective interaction Hamiltonian close to the CDW instability as
\begin{align}
H_{\mathrm{eff}}^{\Lambda_{\mathrm{c}}} = -\frac{1}{\mathcal{N}} \sum_{o,o^{\prime}} V_{o,o^{\prime}}\epsilon_{o}\epsilon_{o^{\prime}} N_{\bm{0}}^{o} \, N_{\bm{0}}^{o^{\prime}},
\label{eq:HCDW}
\end{align}
where $V_{o,o^{\prime}} > 0$, $\mathcal{N}$ is the number of unit cells and  $\epsilon_{A} = +1$, $\epsilon_{B} = -1$ parametrize the sublattice modulation, as previously described in Ref.~\onlinecite{PhysRevB.92.155137}. This approximate effective Hamiltonian factorizes into a sum of products of two density operators with zero momentum transfer, i.e., $N_{\bm{q}}^{o} = \sum_{\bm{k}} c^{\dagger}_{\bm{k}+\bm{q},o}c_{\bm{k},o}$ at $\bm{q}=\bm{0}$ and thus corresponds to a long-ranged density-density interaction, favoring enhanced occupancy on one sublattice and suppressed occupancy on the other. We thus observe that the system is eventually driven toward the commensurate CDW instability that also emerges in the large-$V$ regime at half-filling. This behavior is in fact expected form the observation that in the large-$V$ limit (at $t=0$) the spinless fermion model $H$ maps onto an antiferromagnetic Ising model, 
\begin{align}
H_I= J \sum_{\langle \bm{i}, \bm{j} \rangle} \sigma_{\bm{i}} \sigma_{\bm{j}} - h\sum_{\bm{i}} \sigma_{\bm{i}},
\end{align}
with full (empty) sites represented by $\sigma_{\bm{i}}=+1$ ($\sigma_{\bm{i}}=-1$). Under this mapping, $J=V/4>0$, and  the chemical potential term results in the magnetic field $h=\mu/2$.  As a function of $h$, this Ising model has an antiferromagnetic ground state for $|h|<3J$, whereas it is fully polarized up (down) for $h>3J$ ($h<-3J$). For the spinless fermion model $H$, this implies that in the large-$V$ limit, the lattice is either empty or full, or it is locked into the commensurate CDW phase that is stabilized also at half-filling. As discussed in the following sections, we indeed observe the commensurate CDW to eventually prevail for large values of $V$ throughout the full range of the chemical potential. We finally note, that consistent with this argument, the predominance of the commensurate CDW phase at large interactions was also found in studies of hard-core bosons with nearest-neighbor repulsions on the honeycomb lattice~\cite{PhysRevB.75.174301,PhysRevB.75.214509}.
%
\subsection{Bond-order instability}

%
\begin{figure}[t]
\includegraphics[width=\columnwidth]{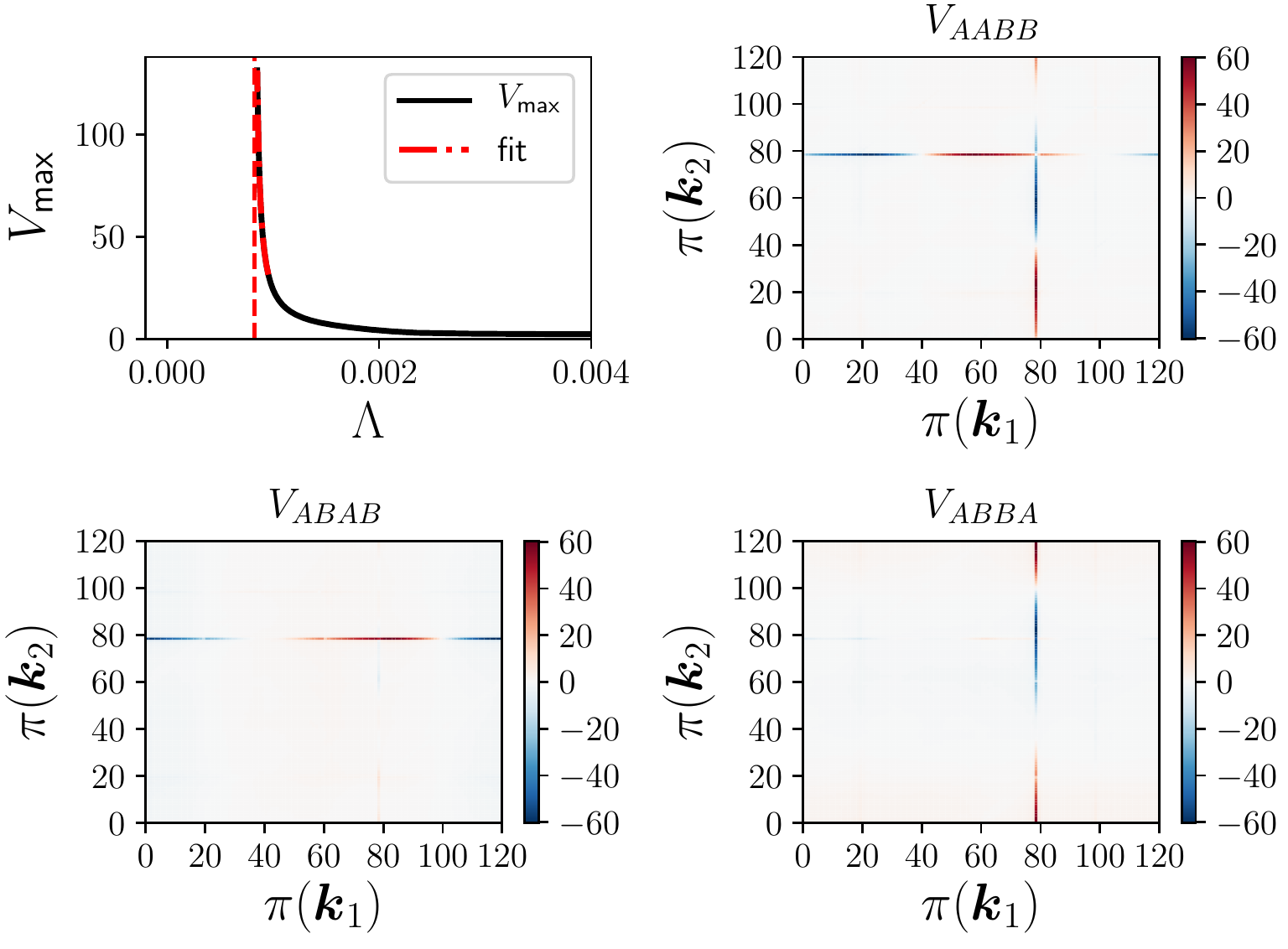}
\caption{(Color online)  Upper left panel: Largest vertex component $V_{\mathrm{max}}$ as a function of the RG scale $\Lambda$ for $\mu/t=-1.0$ and $V/t=0.4$. We identify the critical scale at $\Lambda_c/t\approx 0.0008$.
Upper right panel and bottom panels: Vertex structure for the divergent bond-order correlations for sublattice combinations $V_{AABB}^{\Lambda_{\mathrm{c}}}$, $V_{ABAB}^{\Lambda_{\mathrm{c}}}$, and $V_{ABBA}^{\Lambda_{\mathrm{c}}}$, respectively.
Here, we have chosen $N=120$ patch points following the scheme indicated in Fig.~\ref{fig:patch}~(c) and $\pi(\bm{k}_{3})$ is fixed to the first patch point. 
 The divergent wave-vector structure can be translated to the effective Hamiltonian in Eq.~\eqref{eq:HCBO}.}
\label{fig:vertex_cbo}
\end{figure}
%

At $\mu/t=-1$ the Fermi surface of the tight-binding Hamiltonian is perfectly nested, cf. Figs.~\ref{fig:patch}~(b) and \ref{fig:patch}~(c), and the density of states (DOS) is enhanced due to a Van Hove singularity. Therefore it can be expected that particle-hole fluctuations play a leading role for the possible many-body instabilities. We indeed find an immediate instability of the metallic  phase of the free system for arbitrarily small values of the interaction. The divergent vertex structure is depicted in Fig.~\ref{fig:vertex_cbo} and it exhibits finite momentum transfers of $\bm{q} = \bm{M}_i$, which correspond to the three in-equivalent $\bm{M}$ points in the BZ, see also Fig.~\ref{fig:lattice}. Namely, the transfer momentum between two of the three inequivalent $\bm{M}$ points from the BZ is equivalent to the third $\bm{M}$ point up to a reciprocal lattice vector. For example, $\bm{M}_1-\bm{M}_2 \equiv \bm{M}_3$.
The effective interaction Hamiltonian, capturing a BO instability, can thus be expressed in terms of inter-orbital contributions as
\begin{align}
H_{\mathrm{eff}}^{\Lambda_{\mathrm{c}}} &= -\frac{1}{\mathcal{N}} \sum_{i=1}^{3} V_i\, \chi_{\bm{M}_i}^{\dagger} \chi_{\bm{M}_i}
\label{eq:HCBO}
\end{align}
with
\begin{align}
\chi_{\bm{M}_i} &= \sum_{\bm{k}} \sum_{o} f_{\bm{M}_i}(\bm{k}) c^{\dag}_{\bm{k},o} c_{\bm{k}-\bm{M}_i,\bar{o}},
\end{align}
and where $V_i > 0$, and $f_{\bm{M}_i}(\bm{k})$ is a form factor that we examine next. Namely, to provide a more physical interpretation of Eq.~\eqref{eq:HCBO} and the corresponding instability, we perform a projection of the form factor onto its most relevant components. To this end, we parametrize the divergent vertex components into matrix form,  such that
\begin{align}
V_{ABAB}(\vec{k}, \vec{k}^\prime)= &V^{\Lambda_c}_{ABAB}(\vec{k},\vec{k}^\prime,\vec{k}^\prime-\vec{q},\vec{k}+\vec{q}), \\
V_{ABBA}(\vec{k}, \vec{k}^\prime)=  &V^{\Lambda_c}_{ABBA}(\vec{k},\vec{k}^\prime,\vec{k}+\vec{q},\vec{k}^\prime-\vec{q}),
\end{align}
where $\bm{q}$ is one of the nesting vectors $\bm{M}_i$, and we restrict $\bm{k}$ and $\bm{k}^\prime$ to patches that are connect by $\bm{q}$. In addition, the component $V_{AABB}$ contains contributions from both these parametrizations, as seen explicitly also in   Fig.~\ref{fig:vertex_cbo}. We next perform a numerical eigenmode analysis  of the above matrices using a singular value decomposition. 

\begin{figure}[t]
\vspace{0.25cm}
\includegraphics[width=0.95\columnwidth]{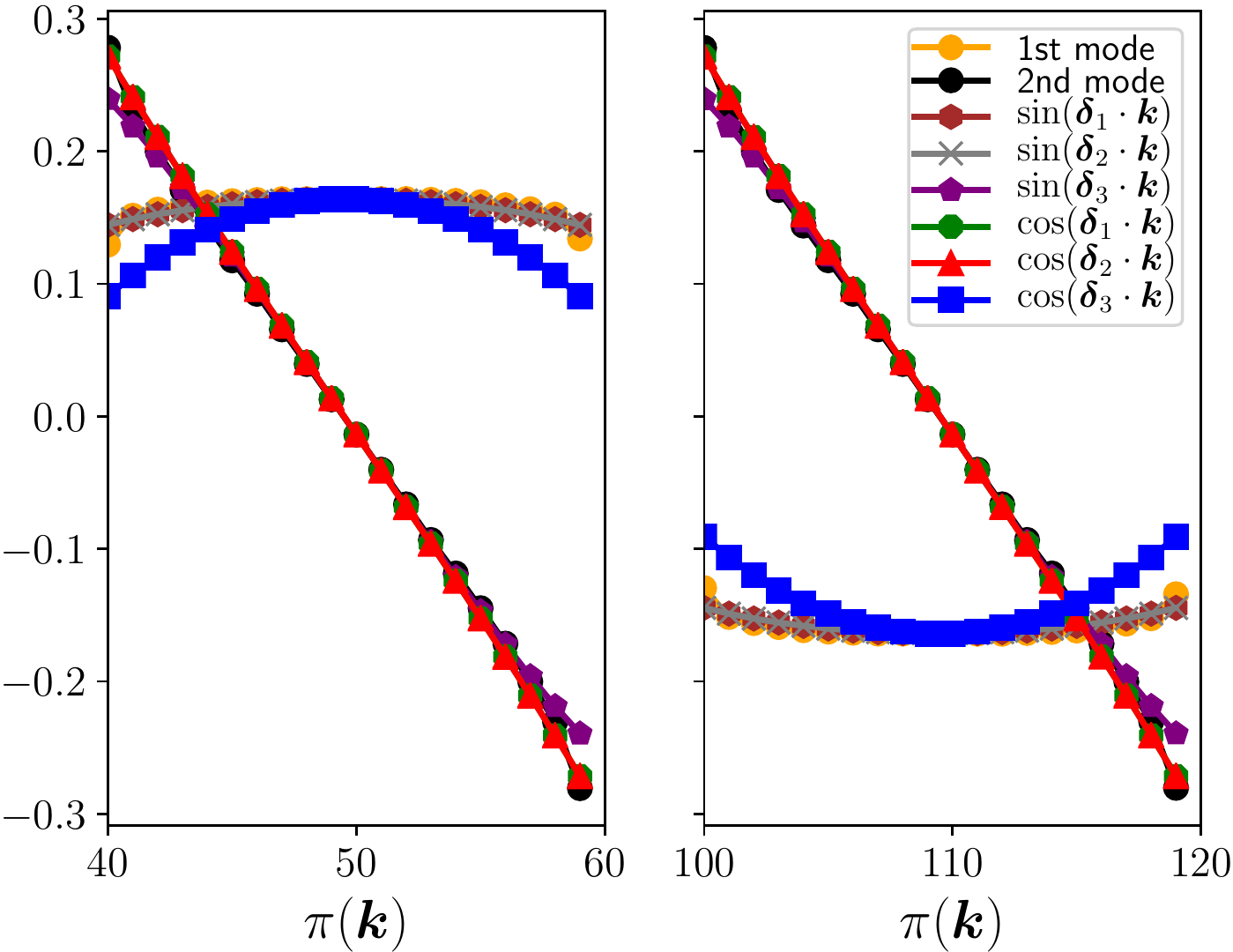}
\caption{(Color online) Comparison of the two dominant eigenmodes of ${V}_{ABBA}(\vec{k}, \vec{k}^\prime)$ with the harmonic BO form factors. The functions are evaluated over points of the Fermi surface connected by the same nesting vector $\bm{q} = \bm{M}_3$, and the 
prefactor of the form factor was scaled to match the eigenmodes at the central momenta.}
\label{fig:eigen_modes_vh}
\end{figure}

We then find that both matrices ${V}_{ABAB}(\vec{k}, \vec{k}^\prime)$ and ${V}_{ABBA}(\vec{k}, \vec{k}^\prime)$ contain  two dominant eigenmodes, while the modulus of other eigenvalues are several orders of magnitude smaller. In particular, the relevant eigenmodes of both ${V}_{ABAB}(\vec{k}, \vec{k}^\prime)$ and ${V}_{ABBA}(\vec{k}, \vec{k}^\prime)$ differ mainly by a sign, related to the exchange of two fermionic operators. As an example, 
we plot  in Fig. \ref{fig:eigen_modes_vh} the resulting eigenmodes  for ${V}_{ABBA}(\vec{k}, \vec{k}^\prime)$  over the connected momentum patches for $\bm{q} = \bm{M}_3$, and compare to   harmonic form factors proportional to 
$\sin(\bm{\delta} \cdot \vec{k})$, and $\cos(\bm{\delta} \cdot \vec{k})$, respectively,
where the vector $\bm{\delta}=\bm{\delta}_i, \, i=1,2,3,$ connects nearest-neighbor bonds as illustrated in Fig. \ref{fig:lattice}. 
We observe a good overall agreement between the extracted eigenmodes and these form factors, which  are characteristic of BO instabilities~\cite{PhysRevB.90.045135}.

%
%

%
\begin{figure}[ht]
\includegraphics[width=\columnwidth]{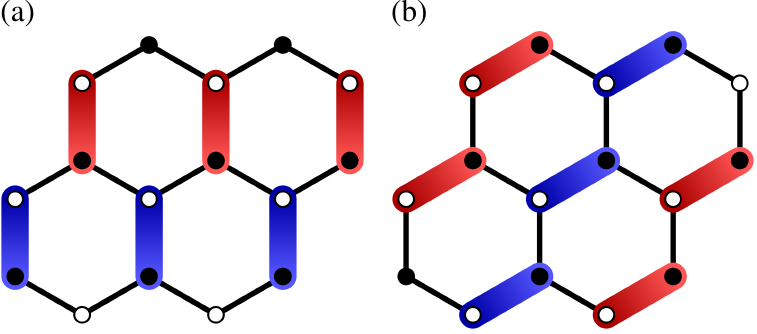}
\caption{(Color online) Schematic real-space hopping renormalization due to a finite bond-order parameter $ \langle \chi_{\vec{M}_{i}} \rangle $. The white and black disks denote the lattice sites of the $A$ and $B$ sublattice, respectively. The bonds shown in red (blue) correspond to an increase (decrease) of the nearest-neighbor hopping amplitude along a given bond (pointing from the $A$ to the $B$ sublattice).
If  $\bm{q} = \bm{M}_i$ is  parallel to  $\bm{\delta}$, the resulting BO pattern 
corresponds to the one shown in panel (a). There are three such states, related by rotations. 
For $\bm{q} = \bm{M}_i$   not parallel to  $\bm{\delta}$,
 a patten such as the one shown in panel (b) results instead. There are a total of six such states, related by rotations or reflections. }
\label{fig:bondorder}
\end{figure}
%

If one performs the Fourier transform of the form factors back to real space, one finds a renormalization of the hopping amplitude with a doubled unit  cell~\cite{PhysRevB.90.045135}, as shown in Fig.~\ref{fig:bondorder}. More specifically, these BO states belong to  two distinct classes, depending on whether $\bm{q} = \bm{M}_i$ is  parallel to  $\bm{\delta}$ or not. 
If  $\bm{q} = \bm{M}_i$ is  parallel to  $\bm{\delta}$, the resulting BO pattern 
exhibits equal amplitudes on parallel dimers across the hexagons,
 such as shown in   Fig.~\ref{fig:bondorder}(a). There are  three such states, related to each other by rotations. 
For $\bm{q} = \bm{M}_i$ not parallel to  $\bm{\delta}$,
 a patten such as the one shown in the Fig.~\ref{fig:bondorder}~(b) results, with zig-zag lines of equal amplitude bonds. There are a total of six such states, related to each other by rotations or reflections.  Within our calculations, we cannot discern which of these two BO classes will be preferably form eventually. 
In any case, 
 the order parameter for a such a BO  state can be interpreted as a translation-symmetry breaking dimerization of the fermionic states on the $A$ and $B$ sublattices.

\subsection{Superconducting $f$-wave instability}

\begin{figure}[t]
\includegraphics[width=\columnwidth]{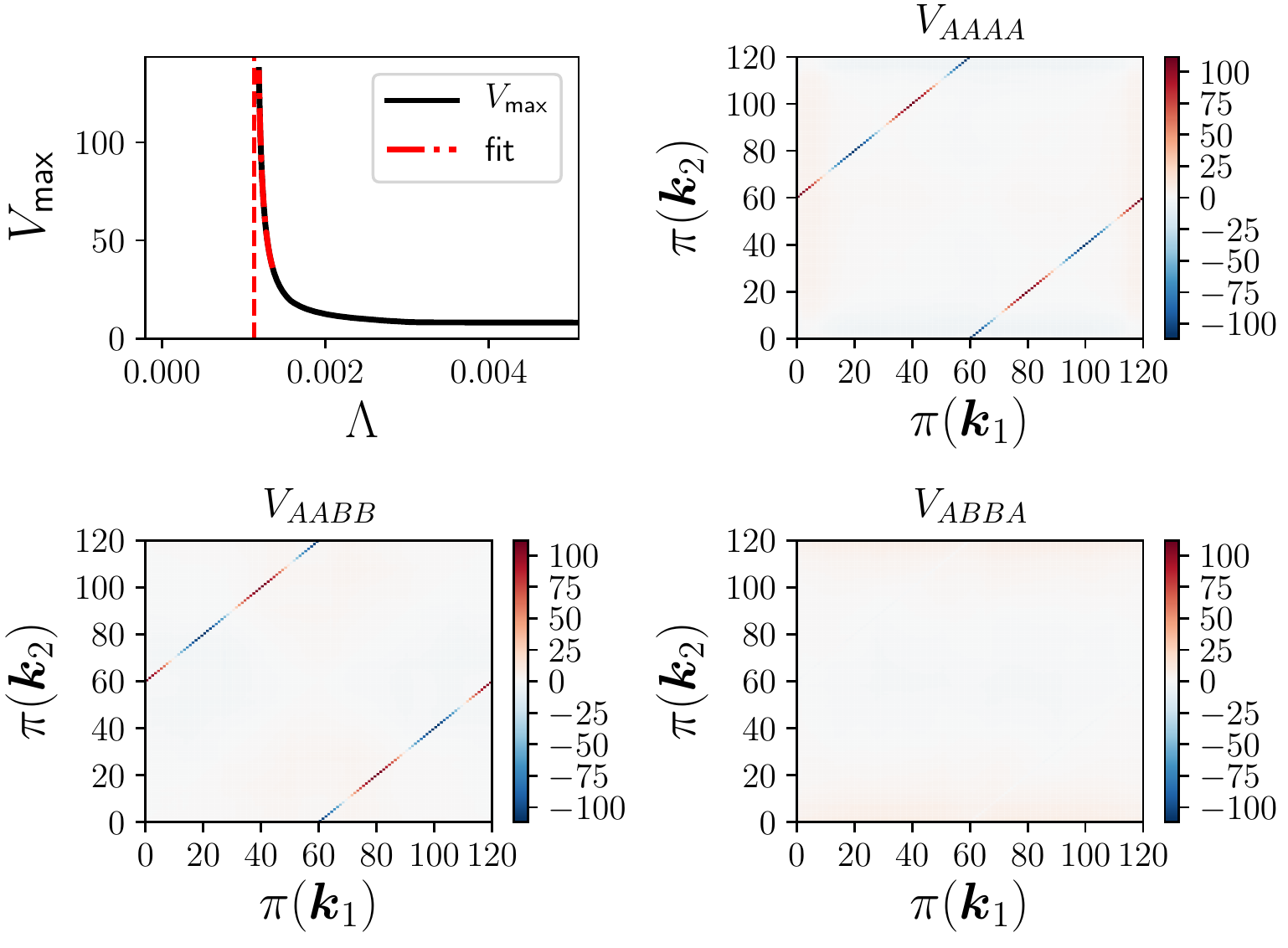}
\caption{(Color online) Upper left panel: Largest vertex component $V_{\mathrm{max}}$ as a function of the RG scale $\Lambda$ for $\mu/t=-1.4$ and $V/t=1.0$. We identify the critical scale at $\Lambda_c / t \approx 0.0011$.
Upper right panel and bottom panels: Vertex structure for the divergent $f$-wave correlations for sublattice combinations $V_{AAAA}^{\Lambda_{\mathrm{c}}}$, $V_{AABB}^{\Lambda_{\mathrm{c}}}$, and $V_{ABBA}^{\Lambda_{\mathrm{c}}}$, respectively. 
Here, we have chosen $N=120$ patch points following the scheme indicated in Fig.~\ref{fig:patch}~(d) and $\pi(\bm{k}_{3})$ is fixed to the first patch point. 
The divergent wave-vector structure can be translated to the effective Hamiltonian in Eq.~\eqref{eq:HSC}.}
\label{fig:vertex_f_wave}
\end{figure}
%

\begin{figure}[t]
\includegraphics[width=\columnwidth]{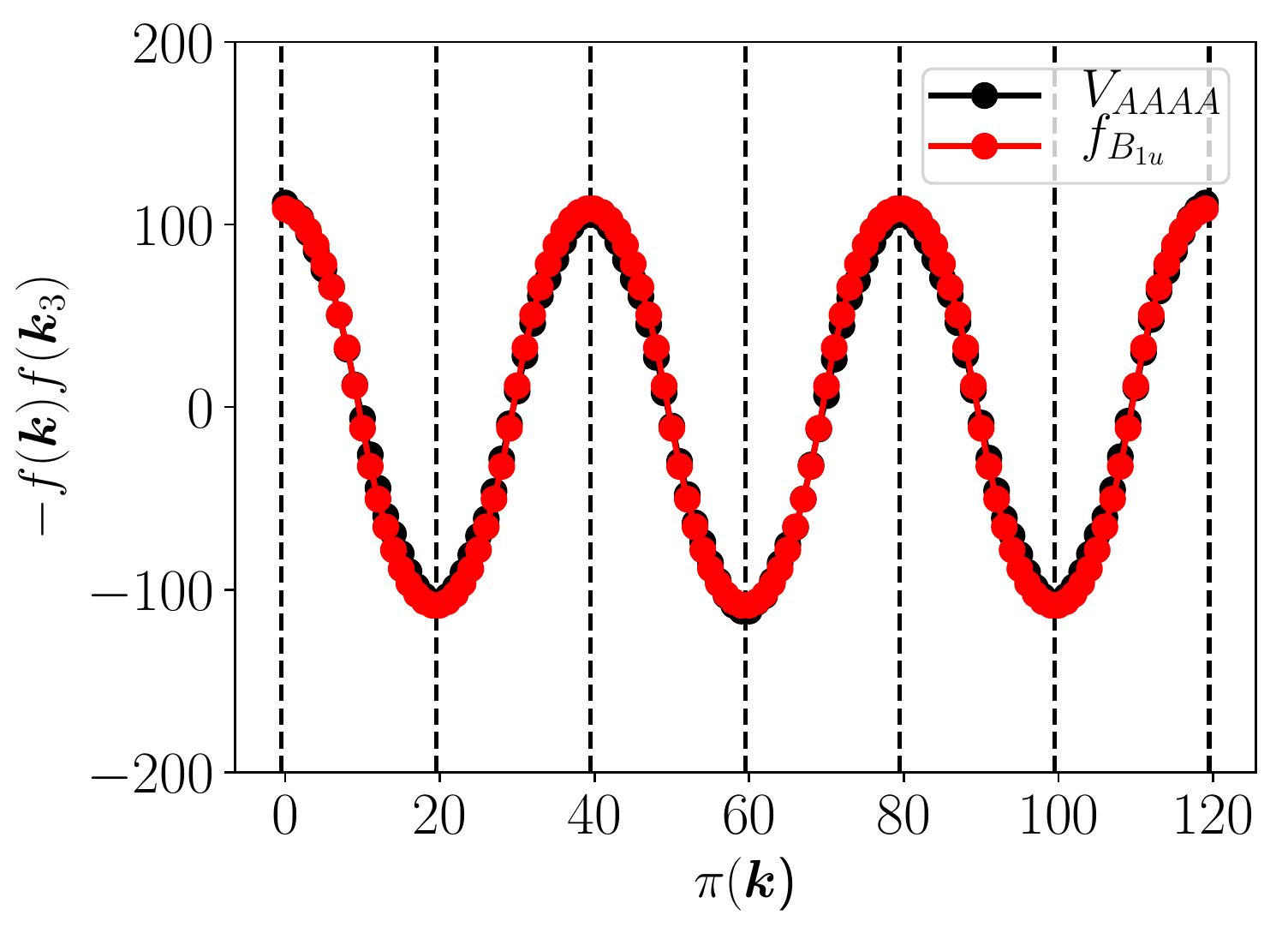}
\includegraphics[width=0.7\columnwidth]{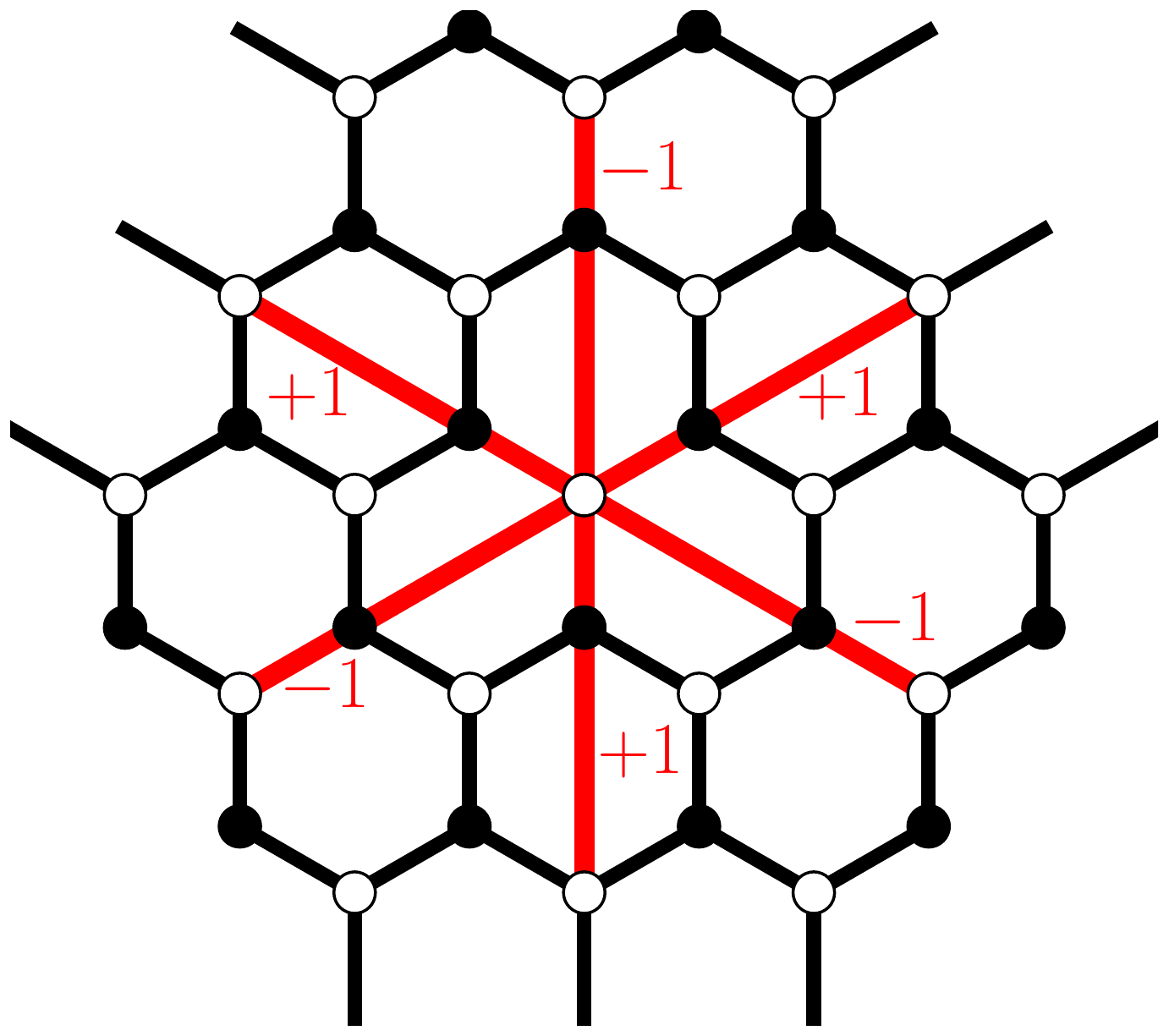}
\caption{ (Color online) Top panel:  Diagonal $V_{AAAA}^{\Lambda_{\mathrm{c}}}$ vertex structure along the Fermi surface. We fix the third momentum to $\pi(\bm{k}_3) = 1$ and perform a fit to the $B_{1u}$ lattice harmonic function $f_{B_{1u}}(\bm{k}) \sim \sin(3 k_y) - 2 \sin(\frac{3}{2} k_y) \cos(\frac{3\sqrt{3}}{2} k_x)$. Bottom panel: Real-space pairing form factor corresponding to the lattice harmonic function $B_{1u}$ on the honeycomb lattice, with the two sublattices $A$ and $B$ indicated  by  blue and green circles, respectively. 
The numbers $\pm 1$ along the (red) lines to the second-nearest equal-sublattice neighbors of  the central site denote  phase factors of the corresponding $B_{1u}$ pairing state.}
\label{fig:diagonal_f_wave}
\end{figure}

For values of the chemical potential beyond the Van Hove singularity, i.e., $\mu/t < -1$, we observe an instability in the particle-particle channel, cf.~Fig.~\ref{fig:vertex_f_wave}. 
This SC instability leads to a characteristic diagonal structure of the vertex, which indicates pairing of fermions with momentum $\bm{k}$ and $-\bm{k}$ and thus signals the formation of Cooper pairs. We therefore deduce the form of the effective Hamiltonian as
\begin{align}
H_{\mathrm{eff}}^{\Lambda_c} &= - \frac{1}{\mathcal{N}} \sum_{o, o^\prime} V_{o o^\prime} \Phi_{o}^{\dagger} \Phi_{o^\prime},
\label{eq:HSC}
\end{align}
with
\begin{align}
\Phi_{o} &= \sum_{\bm{k}} f (\bm{k}) \ c_{-\bm{k},o} c_{\bm{k},o},
\end{align}
and where the symmetry of the SC order parameter is encoded in the form factor $f(\bm{k})$. We illustrate the momentum dependence along the diagonal of $f(\bm{k})$ in the top panel of Fig.~\ref{fig:diagonal_f_wave}. The inter-sublattice components of the vertex show no divergent structures, compatible with the suppression of intra-unit-cell pairing correlations due to the nearest-neighbor repulsion. By the fermionic exchange symmetry, an intrasublattice solution has to be odd with respect to $ \vec{k} \to -\vec{k} $. The sign of the form factor in the intrasublattice channel changes six times at momenta where the $ \Gamma - K$ lines cross the Fermi surface, which is compatible with $f$-wave superconductivity, with a form factor
$f_{B_{1u}}(\bm{k}) \sim \sin(3 k_y) - 2 \sin(\frac{3}{2} k_y) \cos(\frac{3\sqrt{3}}{2} k_x)$
corresponding to the $B_{1u}$  irreducible representation of the $D_{6h}$ symmetry~\cite{BlackSchaffer2014}. 
We note, that on the level of our instability analysis, the relative phase between the SC order parameters on the $A$ and $B$ sublattices is not fixed.

%
%

In real space, such a $B_{1u}$  form factor results in a sign-alternating pairing state on the second-nearest intrasublattice neighbor  bonds, as illustrated by the phase factors shown in the bottom panel of Fig.~\ref{fig:diagonal_f_wave}. 
To rationalize the apparent emergence of such a $f$-wave pairing state in the $t-V$ model on the honeycomb lattice, we remark that 
(i) a nearest-neighbor pairing dominated $p$-wave  state  is  suppressed by
the repulsive nature of the bare repulsive interactions $V>0$ in this model, and 
(ii) nearest intrasublattice neighbor dominated $B_{2u}$ ($f$-wave) pairing may be more prone to a direct pair-breaking transition into the CDW instability than the above $B_{1u}$ pairing state, which features a more spatially extended pair wave-function.

We further note that the pairing instability appears at much smaller critical scales than the particle-hole instabilities (CDW, BO) and is in this respect reminiscent of the fRG studies of the $d$-wave SC instability on the doped Hubbard model on the square lattice with spin-1/2 electrons~\cite{PhysRevLett.87.187004,RevModPhys.84.299}, driven by antiferromagnetic spin-fluctuations. In distinction to the square lattice, however, the honeycomb system has a two-atom unit cell that renders the CDW an instability in the particle-hole channel with vanishing momentum transfer. Correspondingly, in the spinless fermion system on the honeycomb lattice, we can expect the charge fluctuations close to vanishing momentum transfer to mediate an attractive interaction between the fermions. For spin-1/2 fermions, SC $f$-wave triplet solutions have been reported on several types of lattice geometries. There exist suggestions in favor of $f$-wave instabilities generated by onsite repulsion for triangular and/or honeycomb~\cite{doi:10.1143/JPSJ.73.17,PhysRevB.89.144501} lattices. In accordance with the mechanism studied here for the spinless case, proximity to a CDW instability on the triangular~\cite{doi:10.1143/JPSJ.73.2053,doi:10.1143/JPSJ.74.2901,doi:10.1143/JPSJ.73.319}, honeycomb~\cite{PhysRevLett.100.146404}, and also the square~\cite{PhysRevB.70.094523} lattice supports $f$-wave solutions. More recently, spin-orbit coupling~\cite{doi:10.7566/JPSJ.85.104704,doi:10.7566/JPSJ.82.014702} has been suggested to favor SC solutions with $f$-wave symmetry.

\begin{figure}[t]
\includegraphics[width=\columnwidth]{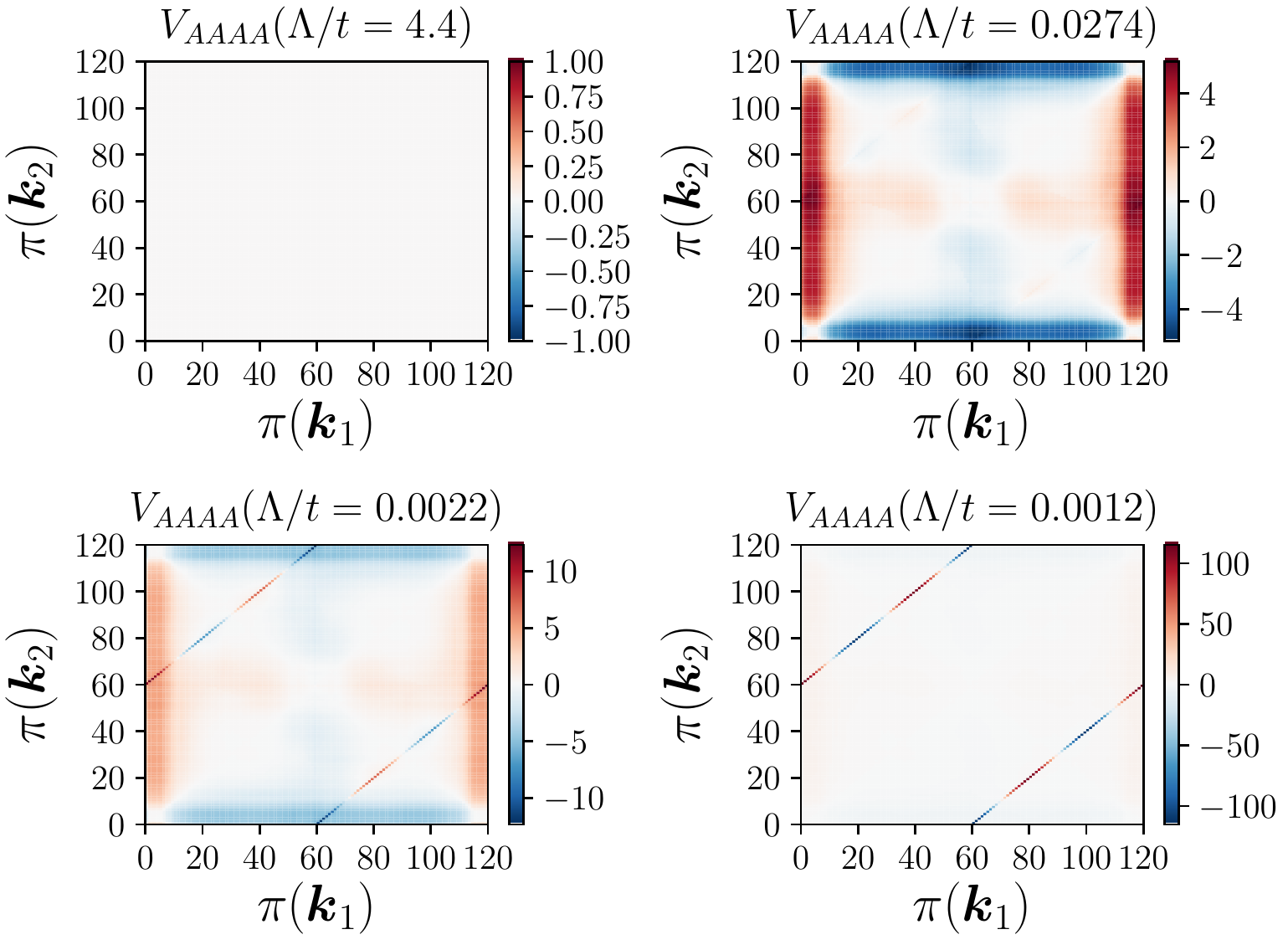}
\caption{(Color online) Evolution of the $V_{AAAA}$ vertex component, shown at various values of the scale $\Lambda$ during the flow. The model parameters are the same as in Fig.~\ref{fig:vertex_f_wave}.}
\label{fig:flow_f_wave}
\end{figure}
%
For the present case, we show the evolution of the vertex function in the parameter regime of the $f$-wave instability in Fig.~\ref{fig:flow_f_wave}.  As the cutoff scale is decreased, we observe that the vertex function first develops features which resemble a CDW, which can be described by a Hamiltonian as in Eq.~\eqref{eq:HCDW} by admitting a finite momentum transfer $ \vec{q} $.
These CDW features create an attractive component in the pairing channel which subsequently grows upon further lowering of the cutoff. Eventually, the pairing channel becomes the leading structure of the vertex function and develops an instability. Furthermore, the phase structure of the pairing state is pinned to the effective intermediate vertex structure, which matches the $\bm{k}$-dependence of the $B_{1u}$ form factor. We note, that while the proximity of the system to a CDW instability and the concomitant charge fluctuations are crucial to obtain a Cooper instability at the observed critical scales, the feedback of the particle-particle channel onto the particle-hole channel are actually essential in suppressing the CDW ordering tendencies to give way to a SC instability. This effect can be demonstrated by excluding the particle-particle bubble $ \Phi_{\mathrm{pp}} $ from the flow Eq. \eqref{eq:flow_eq} and comparing the resulting phase diagram and its critical scales to the one obtained from the flow with the particle-particle bubble included. For the region where we observe a SC $f$-wave instability, we find the critical scale for the CDW instability with the particle-particle bubble excluded to always be above the critical scale for the SC instability. For chemical potentials with $ | \mu | < t $, the suppression of the CDW ordering tendencies is not strong enough to allow for the SC correlations to become leading. 
Finally, let us note that in the semimetallic region of the phase diagram, we can exclude a Kohn-Luttinger instability down to critical scales $ \Lambda/t = 10^{-7}$, where we stopped the integration of an otherwise regular flow.

\section{Conclusions}\label{sec:conclusions}

We have studied the many-body instabilities of spinless fermions on the honeycomb lattice beyond half-filling, employing the fRG in the Fermi-surface patching scheme.
Our work extends on previous studies of the half-filled case, which has been intensely explored with many different many-body methods~\cite{PhysRevLett.100.156401,PhysRevB.81.085105,PhysRevLett.107.106402,PhysRevB.87.085136,PhysRevB.92.155137,PhysRevB.92.085146,PhysRevB.88.245123,PhysRevB.89.035103,PhysRevB.89.165123,PhysRevB.92.085147}, including the fRG~\cite{PhysRevB.92.155137}.
For half-filling and with large enough nearest-neighbor repulsion, the different approaches consensually find a CDW ordering transition and convergence in the critical interaction as well as the critical behavior is currently being established.
Beyond half-filling, on the other hand, only a few results are available thus far. 
In particular, the mean-field approach in Refs.~\onlinecite{PhysRevLett.107.106402, PhysRevB.87.085136} left out the possibility of SC order. The approximate numerical approaches 
from Refs.~\onlinecite{PhysRevB.94.075144, Broecker2017} exclusively focus on one-third filling, where only indirect indications for  a quantum phase transition were reported. 
Our functional RG approach supplements these previous studies by a systematic investigation of the system's many-body instabilities over a broad range of filling and coupling strength, taking into account the different competing channels on equal footing.

We have summarized our results in the tentative phase diagram in the $V$--$\mu$--plane; see Fig.~\ref{fig:phase_diagram}.
Between half-filling and the Van Hove singularity points, which correspond to fillings of $3/8$ and $5/8$ in the non-interacting model, a conventional CDW  is found to be the leading instability, once a filling-dependent critical interaction strength is exceeded, sharing its characteristics with the one of the half-filled case.
At the Van Hove singularity point, where the free Fermi surface is perfectly nested and the DOS is strongly enhanced, particle-hole fluctuations induce a dimerized bond-order instability, which appears even for arbitrarily small values of the interaction.
For larger interactions, the BO instability is superseded by the CDW  instability.
Finally, for chemical potentials beyond the Van Hove point, i.e., $|\mu|>t$, an $f$-wave SC instability emerges.
This pairing instability results from a particle-hole fluctuation-induced attractive component in the intrasublattice pairing channel.
After the attractive interaction has been created, it eventually becomes the leading instability upon integrating out the renormalization group flow.
Due to this two-step process, the associated critical scales of the pairing instability are considerably smaller than the ones from the plain particle-hole instabilities, i.e., the CDW and the bond ordering. Upon further increasing $V$, the system again enters the commensurate CDW state. This result is in accord with the large-$V$ (Ising model) limit of this model.  While we cannot discern the nature of the quantum phase transition between the pairing phase and the CDW regime, it is expected to be discontinuous, based on the distinct symmetries that are broken within these two phases.  Correspondingly, in such a scenario, the large-$V$ regime is characterized by phase separation in the canonical ensemble within the corresponding density regime, similarly to the related hard-core boson model~\cite{PhysRevB.75.174301,PhysRevB.75.214509}.

In conclusion, our tentative phase diagram of the spinless fermion $t-V$ model 
on the honeycomb lattice, which we obtained from the Fermi-surface patching weak-coupling functional RG approach, connects well to exactly known or well-established results in several limiting cases, including the strong coupling limit. 

The availability of known results in various limits of the $t-V$ model  facilitates the assessment of the validity of our findings.
This contrasts  to systems with several  flavors of fermions, for which the case of spin-1/2 fermions on the honeycomb lattice assuredly is
the most prominent one in view of, e.g., the physics of graphene. In the spinful case, the possibility for many-body instabilities is significantly enriched  by the fluctuations in the spin degree of freedom and their interplay with charge and pairing correlations. Moreover, in a tight-binding description, a local on-site (Hubbard $U$) repulsion is expected to be the dominant interaction term and  this affects the leading many-body instabilities. This is the case already  at half-filling, for which  a commensurate spin density wave (SDW), i.e., a two-sublattice antiferromagnetic instability, emerges for dominant onsite repulsion~\cite{tchougreeff1992charge,0295-5075-19-8-007,PhysRevLett.97.146401,meng2010quantum,sorella2012absence,assaad2013pinning}. To stabilize a commensurate CDW corresponding to the one that we obtained for the spinless $t-V$ model of spinless fermions, one  thus requires a suppression of the SDW instability, e.g., by a sufficiently enhanced nearest-neighbor repulsion, alike $V$ in the spinless model~\cite{tchougreeff1992charge,PhysRevLett.97.146401,PhysRevLett.100.156401,volpez2016electronic,de2017competing}.
Doping beyond half-filling, singlet SC instabilities, with a symmetric orbital sector, can emerge in the spinful system, whereas  pairing states with a  symmetric  orbital structure are not possible for spinless fermions. In various recent  works on SC instabilities of spin-1/2 fermions on the honeycomb lattice, a chiral $d$-wave pairing state has been identified as the most dominant pairing channel over different ranges of doping, see Ref.~\onlinecite{BlackSchaffer2014} for a review. Close to the Van Hove singularity, further instabilities were found to strongly compete with  the $d$-wave pairing. Among these, $f$-wave pairing and a chiral SDW state have been reported near the Van Hove filling~\cite{Li2012,PhysRevB.86.020507,PhysRevB.85.035414,Ying2018}. 
In addition, Ref.~\onlinecite{PhysRevB.90.045135} reports BO instabilities for a spin-1/2 Kitaev-Heisenberg model doped to the Van Hove filling, resembling those that we identified in the case of the spinless fermion model.  Thus, spin fluctuations provide a rather rich variety of additional and competing many-body instabilities for spin-1/2 fermions on the honeycomb lattice, while the reduced complexity of the spinless fermion model that we considered here, allows us to provide   a consistent phase-diagram based on  the functional RG approach.

With regards  to graphene, recent experiments on twisted bilayer graphene which---depending on the filling level---find strongly correlated insulating behavior~\cite{Jarillo-Herrero:2018a} and superconductivity~\cite{Jarillo-Herrero:2018} have spurred ample excitement.
Currently, many theoretical models for twisted honeycomb bilayers with electronic correlations are conceived, see, e.g., Refs.~\onlinecite{XuBalents2018,PhysRevB.98.045103,2018arXiv180309742P,PhysRevB.97.235453,2018arXiv180403162D}, but given the juvenile experimental situation it is difficult to judge the validity of any of these suggestions.
Regardless, an improved theoretical understanding of correlated fermions on honeycomb lattice structures seems mandatory. 
Here, we have taken the approach of studying the doped single-layer system with spinless fermions as a basic building block which can define a starting point for the investigation of more complex honeycomb structures in the future.

\section*{Acknowledgments}

We thank Carsten Honerkamp, Giulio Schober, Thomas~C.~Lang and Lei Wang for discussions, and acknowledge support by the Deutsche Forschungsgemeinschaft (DFG) under Grants No. FOR 1807 and No. RTG 1995. Furthermore, we thank the IT Center at RWTH Aachen University and the JSC J\"ulich for access to computing time through JARA-HPC. D.D.S. acknowledges support from the Carlsberg Foundation. M.M.S. was supported by the DFG through the Collaborative Research Center SFB1238, TP C04.
%

\bibliography{spinlesshoney}

\end{document}